\documentclass[letterpaper, 10 pt, conference]{ieeeconf}  
\usepackage{graphicx}
\usepackage{amsfonts}
\usepackage{amsmath}
\usepackage[colorlinks=true,linkcolor=blue]{hyperref}
\usepackage{algorithm}
\usepackage{algorithmicx}
\usepackage{amsmath}
\usepackage{amssymb}
\usepackage{subcaption}

\usepackage{amsthm}
\newtheorem{lemma}{Lemma}
\newtheorem{theorem}{Theorem}
\newtheorem{remark}{Remark}

\usepackage{ifthen}
\newboolean{showcomments}
\setboolean{showcomments}{true}
\usepackage[usenames,dvipsnames]{color}

\definecolor{bleudefrance}{rgb}{0.19, 0.55, 0.91}
\definecolor{ao(english)}{rgb}{0.0, 0.5, 0.0}

\newcommand{\addcite}[0]{\ifthenelse{\boolean{showcomments}}
{\textcolor{purple}{(add cite(s)) }}{}}%

\newcommand{\you}[1]{  \ifthenelse{\boolean{showcomments}}
{{\color{blue} PCY: #1}}{}}
\newcommand{\youmargin}[1]{\ifthenelse{\boolean{showcomments}}{\marginpar{\color{bleudefrance}\tiny PCY: #1}}{}}

\newboolean{showedits}
\setboolean{showedits}{true}
\usepackage[markup=underlined]{changes}
\definechangesauthor[color=bleudefrance]{PCY}
\newcommand{\ayou}[1]{
\ifthenelse{\boolean{showedits}}
{\added[id=PCY]{#1}}
{\!#1\hspace{-4.75pt}}
}
\newcommand{\ryou}[2]{
\ifthenelse{\boolean{showedits}}
{\replaced[id=PCY]{#1}{#2}}
{\!#1\hspace{-4.75pt}}
}
\newcommand{\dyou}[1]{
\ifthenelse{\boolean{showedits}}
{\deleted[id=PCY]{#1}}
{}
}
\newcommand{\hl}[1]{\ifthenelse{\boolean{showcomments}}
{\textcolor{purple}{#1}}{}}%

\IEEEoverridecommandlockouts                              
\title{\LARGE \bf
Online Electricity Purchase for Data Center with Dynamic Virtual Battery from Flexibility Aggregation
}

\author{Kekun Gao, Yuejun Yan, Yixuan Liu, Endong Liu, and Pengcheng You 
\thanks{This work was supported by NSFC through grants 72201007, T2121002, 72131001, as well as CCF-Alibaba Innovative Research Fund.}
 \thanks{K. Gao, Y. Liu, and P. You are with the Department of Industrial Engineering and Management, College of Engineering, Peking University, Beijing, China.
        {\tt\small pcyou@pku.edu.cn}}%
\thanks{Y. Yan is with Alibaba Group, Hangzhou, China.}%
\thanks{E. Liu is with the School of Electrical and Mechanical Engineering,
Heze University, Heze, China.}%
}
\begin{document}

\maketitle
\thispagestyle{plain}
\pagestyle{plain}

\begin{abstract}
As a critical component of modern infrastructure, data centers account for a huge amount of power consumption and greenhouse gas emission. This paper studies the electricity purchase strategy for a data center to lower its energy cost while integrating local renewable generation under uncertainty.
To facilitate efficient and scalable decision-making, we propose a two-layer hierarchy where the lower layer consists of the operation of all electrical equipment in the data center and the upper layer determines the procurement and dispatch of electricity.
At the lower layer, instead of device-level scheduling in real time, we propose to exploit the inherent flexibility in demand, such as thermostatically controlled loads and flexible computing tasks, and aggregate them into virtual batteries. By this means, the upper-layer decision only needs to take into account these virtual batteries, the size of which is generally small and independent of the data center scale.
We further propose an online algorithm based on Lyapunov optimization to purchase electricity from the grid with a manageable energy cost, even though the prices, renewable availability, and battery specifications are uncertain and dynamic. In particular, we show that, under mild conditions, our algorithm can achieve bounded loss compared with the offline optimal cost, while strictly respecting battery operational constraints. Extensive simulation studies validate the theoretical analysis and illustrate the tradeoff between optimality and conservativeness.
\end{abstract}

\section{Introduction}
Data centers, serving as the crucial infrastructure within the contemporary digital economy, are experiencing a rapid increase in energy demand owing to their continuously expanding scale.
As large energy consumers, data centers spend nearly 30\% to 50\% percent of their operating costs on energy bills, where a significant proportion of the electricity consumed originates from the electricity spot market.
Due to the huge energy consumption of data centers, a 1\% improvement in power efficiency would save millions of dollars for the cloud service providers, which has led to growing attention on minimizing energy costs through optimizing spot market transactions.
To address this issue, one feasible option is to use the flexibility of data centers' power equipments to reduce energy costs and absorb green power generation.
Currently, data centers are commonly equipped with renewable energy generation, yet their energy supply is highly influenced by climatic conditions, rendering it intermittent and difficult to predict \cite{zhou2017optimal,10136373}. Furthermore, electricity prices in the spot market fluctuate in real time, and the demand-side flexibility of data centers has great randomness. Faced with the randomness of flexible loads, volatility of electricity prices, and intermittency of renewable energy, minimizing the data center in spot market energy cost in the absence of future statistical knowledge poses a challenging problem.

There have been numerous related works addressing the optimization of electricity procurement for data centers.
Some existing works use storage and price changes to address time-dimensional shifts in energy demand \cite{toosi2015fuzzy}, \cite{chen2017stochastic}. Some works formulate these problems as a stochastic optimization problem and introduce Lyapunov optimization technologies to make real-time decisions. \cite{neely2010efficient} and \cite{wei2017data} take the uncertainty of electricity price and load into consideration during optimization, a two-time scale power market is considered in \cite{yao2012power}, and Zhang \emph{et al.} investigate the problem of energy management for data centers with renewable resources and energy storages to minimize electricity cost by leveraging the diversity of these system states \cite{zhang2020distributed}. However, these works either fail to consider energy storage or only consider real battery-based energy storage. In other words, they do not aggregate other flexible loads and energy storage available in data centers to form virtual batteries, resulting in under-utilization of the flexibility of data center loads.

In reality, a data center comprises various heterogeneous classes of flexible loads. We can integrate these loads in the data center into system-level operation and control them by constructing a simple and user-friendly model. Among various modeling methods, the virtual battery model gains its popularity due to its simple and compact form. It aims to quantify the aggregate flexibility while considering the physical constraints of each individual load. Specifically, the virtual battery model constitutes a scalar linear system, akin to a simplified representation of battery dynamics, with parameters denoting charge and discharge power constraints, and energy capacity limits. 
Studies have demonstrated the ability of virtual battery models to effectively capture the aggregate flexibility of diverse classes of flexible loads. \cite{nayyar2013aggregate}, \cite{hao2014aggregate}, and \cite{hao2014characterizing} model the aggregate flexibility of deferrable tasks and Thermostatically Controlled Loads (TCLs), respectively. 
Hao \emph{et al.} first consider four types of flexible loads: TCLs, distributed energy storages, residential pool pumps, and electric vehicles, 
and for each type of load, they derive a closed-form approximation for the aggregate flexibility set based on the individual load parameters \cite{hao2015generalized}. 
\cite{madjidian2017energy} further considers the heterogeneous arrival time of deferrable loads, and also tries to get the close-form battery approximation. In \cite{zhao2017geometric} and \cite{taha2022efficient},  the authors simplify the problem of identifying battery parameters to a polynomial complexity, which further contributes to the application of the virtual battery in the distributed energy resources' flexibility management. Therefore, aggregating the individual flexible loads of the data center as a virtual battery can significantly improve the rationality of electricity purchase and reduce energy costs. 

The objective of this paper is to devise an online strategy for managing the power purchasing policy of data centers that integrate time-varying virtual batteries and renewable energy resources, with the aim of minimizing long-term costs while meeting the energy demand. By leveraging Lyapunov optimization techniques, we develop an online algorithm to address this problem. We demonstrate that the proposed algorithm achieves close-to-optimal performance even in the presence of time-varying parameters of the virtual battery, without requiring prior knowledge of statistics. The cost incurred is at most higher than the optimal algorithm by $O(1/V)$, where $V$ is a controllable parameter balancing the trade-off between energy cost and virtual battery capacity. Furthermore, we provide explicit solutions for control actions under current system dynamics, ensuring the efficiency of our algorithm. Finally, the simulation results validate the effectiveness of the proposed algorithm.

\section{System Model}
\begin{figure}[htbp]
    \centering
    \includegraphics[scale=0.33]{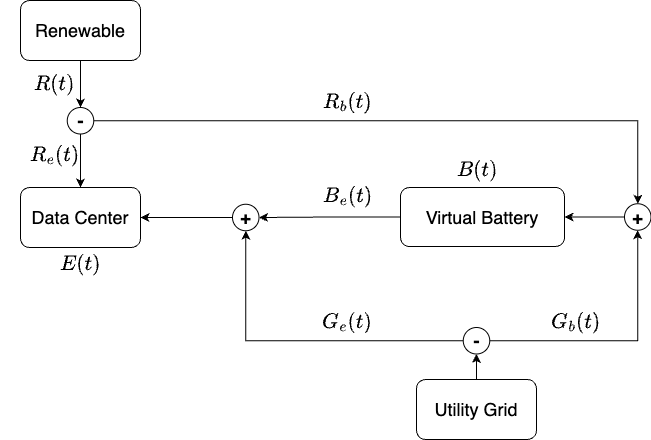}
    \caption{Electricity flows in a data center with virtual batteries.}
    \label{f1}
\end{figure}

Consider the system operation of a data center in a discrete-time horizon $t$ $\in$ \{0, 1, 2,...\}, which uses electricity power from a utility grid and local renewable generation to meet its energy demand. 
We divide the overall demand into two categories: one is \emph{controllable}, e.g., TCLs of cooling systems and flexible computing tasks of IT equipment, and the other is \emph{non-controllable} including necessary power to support basic services, uninterruptible computing tasks with priority, etc.
We propose to aggregate controllable demand into virtual batteries such that the flexibility can be exploited in a computationally efficient way to manage electricity purchase.
A schematic diagram of the system model is given in Figure~\ref{f1}.
We develop the detailed model for each component of the system below.

\subsection{Virtual Battery}

In this subsection, we use TCLs and flexible computing tasks, the two most common loads in the data center, as examples to show how virtual batteries can be aggregated. In particular, a virtual battery is specified by a tuple of parameters $\left(B_{char}(t),B_{dis}(t), B_{min}(t), B_{max}(t), \alpha \right)$, where $B_{char}(t)$/$B_{dis}(t)$ is the charge/discharge rate limit, $B_{min}(t)$/$B_{max}(t)$ is the lower/upper bound of battery capacity, and $\alpha$ is the dissipation rate.
Given such a specification, a virtual battery is modeled by two variables - the battery State-of-Charge (SoC) $B(t)$ and the energy injection/withdrawal $U(t)$ - subject to the following:
\begin{subequations}\label{eq:vb}
\begin{equation} \label{vb1}
-B_{dis}(t) \leq U(t) \leq B_{char}(t), \quad \forall t, 
\end{equation}
\begin{equation}\label{vb2}
    B(t+1)=\alpha B({t})+U({t}),\quad \forall t,
\end{equation}
\begin{equation}\label{vb3}
    B_{min}(t) \leq B({t} )\leq B_{max}(t),  \quad \forall t.
\end{equation}
\end{subequations}



\emph{TCL}: 
The role of a TCL, mainly a cooling system, of the data center is to maintain required room temperatures for IT equipment. Such a requirement is typically captured by a set of thermodynamics equations and operational constraints:
\begin{subequations}
\begin{equation}
    \theta(t+1)=\alpha \theta(t)+\left(1-\alpha \right)\left(\theta_{a}(t) + cr(t)-b p(t)\right), \quad \forall t,
\end{equation}
\begin{equation}
 \theta_r - \Delta \le \theta(t)  \le  \theta_r + \Delta , \quad \forall t,
\end{equation}
\begin{equation}
    0 \le p(t) \le p_m, \quad \forall t.
\end{equation}
\end{subequations}%
Here the variables are the room temperature $\theta(t)$ and the operating power $p(t)$ of the TCL. The external inputs include the operating power of IT devices in the room $r(t)$ and the ambient temperature $\theta_{a}(t)$. The parameters include the room temperature setpoint $\theta_r$ with a tolerance gap $\Delta$, the maximum TCL power $p_m$, the coefficient $c$ that reflects the temperature rise per unit of IT power, and the coefficient $b$ that reflects the temperature drop per unit of the TCL power~\cite{hao2015generalized}, \cite{fang_thermal-aware_2017}.


Note that there exists a nominal value $p_o(t)$ for $p(t)$ that can set the immediate room temperature to the setpoint $\theta_r$, given by 
$$ p_{0}(t)=\frac{\theta_{a}(t) +cr(t)-\theta_{r}}{b}.$$ 
By defining 
$$B(t):=\frac{\theta_{r}-\theta(t)}{(1-\alpha)b}, \quad U(t):=p(t)-p_{o}(t), $$
we obtain 
\begin{subequations}\label{eq:TCL_virtual_battery}
\begin{equation}\label{vb4}
     -p_{0}(t) \le U(t) \le p_{m}-p_{0}(t), \quad \forall t,
\end{equation}
\begin{equation}\label{vb5}
    B(t+1)=\alpha B(t)+U(t), \quad \forall t,
\end{equation}
\begin{equation}\label{vb6}
    -\frac{\Delta}{(1-\alpha)b} \le B(t) \le  \frac{\Delta}{(1-\alpha)b}, \quad \forall t.
\end{equation}
\end{subequations}%
It is obvious from \eqref{eq:TCL_virtual_battery} that the flexibility of a TCL can be represented as a virtual battery, parameterized by
$$\left(p_m-p_o(t),\ p_o(t),\ -\frac{\Delta}{(1-\alpha)b},\ \frac{\Delta}{(1-\alpha)b},\ \alpha\right).$$

\emph{Delay-Tolerant Computing Task}: 
A lot of SQL or machine learning tasks are flexible with the processing schedule, as long as they are completed before a specified deadline. 
Consider a set of such tasks, each (task $j$) parameterized by $(a^j,d^j,\bar L^j, E^j)$, denoting the arrival time, the deadline, the power consumption corresponding to the maximum processing speed, and the total energy required to complete the task, respectively. Let $L^j(t)$ be the allocated power to task $j$ at time $t$. 
Then the task can be characterized as
\begin{subequations}    
\begin{equation}\label{com1}
    0 \leq L^j(t) \leq \bar{L}^j,\quad \forall t \in [a^j,d^j)
\end{equation}
\begin{equation}\label{com2}
    L^j(t)=0,\quad \forall t \notin [a^j,d^j)
\end{equation}
\begin{equation}
    \sum_{t}L^j(t)  = E^j.
\end{equation}
\end{subequations}

The aggregate flexibility of this set of tasks can also be approximated as a virtual battery \cite{hao2015generalized}. Define in this case 
$$B(t):= \sum_{\tau < t}\sum_{j}L^j(t), \quad U(t):= \sum_j L^j(t).$$
Obviously, $U(t)$ cannot exceed the sum of the maximum power consumption of all active tasks, i.e.,
\begin{equation}
   0 \le U(t) \le \sum_{j: a^{j} \leq t < d^{j}} \bar{L}^j.
\end{equation}
Meanwhile, to guarantee that all active tasks at time $t$ can be completed by their deadlines, the SoC $B(t)$ has to achieve a minimum level given by 
\begin{equation}
    B(t)\ge \sum_{j: d^{j} \le t} E^{j}+\sum_{j: a^{j} \leq t < d^{j}} \max \left\{E^{j}-\left(d^{j}-t\right) \bar{L}^j, \ 0\right\} , \nonumber
\end{equation}
where we have accounted for the energy delivered for all finished tasks up until time $t$. Note that the lower bound implies that each active task $j$ will be allocated the maximum processing power from now on and can receive at most $(d^j-t)\bar{L}^j$ amount of energy. 
Similarly, we can derive an upper bound for the SoC $B(t)$ as
\begin{equation}
    B(t) \le \sum_{j: d^{j} \le t} E^{j}+\sum_{j: a^{j} \leq t < d^{j}} \min \left\{E^{j},\left(t-a^{j}\right) \bar{L}^j \right\} . \nonumber
\end{equation}
From above, the parameters of this aggregated virtual battery are available with a dissipation rate $\alpha= 1$. 

\begin{remark}
    As shown from the two virtual battery examples, the parameterization is likely time-varying and uncertain, depending on random ambient temperatures and computing task arrivals, etc. Therefore, managing such dynamic virtual batteries from flexibility aggregation is more challenging than operating real batteries.
\end{remark}

Note that multiple batteries can be readily aggregated into one \cite{hao2015generalized}. Without loss of generality, we consider only one virtual battery with no dissipation ($\alpha=1$) in this work for ease of presentation. The analysis and results also generalize to the concurrent management of multiple virtual batteries. Specifically, in our setting, electricity from both local renewable generation and the utility grid can be used to charge the virtual battery. 
We use $R_b(t)$ and $G_b(t)$ to denote respectively the electricity from the two sources.
Besides, we define $B_e(t)$ to represent the energy withdrawal from the virtual battery to serve the non-controllable demand in the data center. Therefore, the characterization \eqref{eq:vb} of the virtual battery can be explicitly written as
\begin{subequations}\label{VB}
\begin{equation} \label{V1}
0 \leq G_b(t)+R_b(t) \leq B_{char}(t), \quad \forall t,
\end{equation} 
\begin{equation}\label{V2}
   G_b(t)\ge0 , \quad R_b(t)\ge 0, \quad \forall t,
\end{equation}
\begin{equation} \label{V3}
0 \leq B_e(t) \leq B_{dis}(t), \quad \forall t,
\end{equation}
\begin{equation} \label{V4}
B(t+1) =  B(t) + G_b(t) + R_b(t) - B_e(t), \quad \forall t,
\end{equation}
\begin{equation}\label{V5}
        B_{min}(t) \leq B({t} )\leq B_{max}(t),  \quad \forall t.
\end{equation}
\end{subequations}

Furthermore, simultaneous charge and discharge are not allowed:
\begin{equation} \label{V7}
G_b(t)+R_b(t) > 0  \perp B_e(t) > 0 
\end{equation}

\subsection{Problem Formulation}

Suppose the data center is equipped with local renewable generators, such as roof-top solar panels, which can provide an $R(t)$ amount of renewable energy at time $t$. $R(t)$ is random in real time, but will always be upper bounded by solar panels' inverter capacity $R_{max}$, i.e.,
\begin{equation*} \label{R1}
0 \leq R(t) \leq R_{max} .
\end{equation*}
The electricity generated from renewables is divided into three parts: $R_e(t)$ is directly supplied to the data center to meet its (non-controllable ) demand; $R_b(t)$ can be stored in the virtual battery whenever necessary; the rest is just abandoned. This is captured by
\begin{subequations}\label{eq:renewable}
\begin{equation} \label{R2}
0 \leq R_e(t)+R_b(t) \leq R(t) ,\quad \forall t,
\end{equation}
\begin{equation} \label{R3}
R_e(t) \geq 0, \quad \forall t.
\end{equation}
\end{subequations}

Denote the (non-controllable) demand of the data center at time $t$ as $E(t)$. In total, it is satisfied by a combination of electricity supply from the local renewable generation, the virtual battery, and the utility grid. We assume the utility grid is able to offer any amount of electricity that the data center requests to purchase, which yields
\begin{equation} \label{E1}
E(t) = R_e(t)+B_e(t)+G_e(t) , \quad \forall t,
\end{equation}
where $G_e(t)$ denotes the supply from the utility grid that directly serves the energy demand. Meanwhile, the data center can also purchase electricity $G_b(t)$ from the utility grid to store in the virtual battery.
Note that selling electricity back to the utility grid is temporarily not allowed, i.e.,
\begin{equation} \label{C2}
G_e(t) \geq 0, \quad \forall t.
\end{equation}

Let $P(t)\ge 0$ denote the unit price of electricity at time $t$. Then the corresponding energy cost for electricity purchase is
\begin{equation} \label{C3}
C(t) = P(t)[G_e(t)+G_b(t)] .
\end{equation}
The offline problem of the data center over a long (infinite) time horizon can be formulated to minimize the time-average energy cost while satisfying the energy demand at each time~$t$:
\begin{equation} \label{P1}
\textbf{P1: }\min \lim \limits_{T \to \infty} \frac{1}{T}\sum_{t=0}^{T-1} \mathbb{E} \{C(t) \}
\end{equation}
\begin{center}
    \text{s.t.} \eqref{VB},\eqref{V7},\eqref{eq:renewable},\eqref{E1},\eqref{C2}
\end{center}
However, in practice, the electricity prices, renewable generation, and energy demand of data centers are all random but will be sequentially revealed online. Upon observing the current price $P(t)$, renewable availability $R(t)$, and energy demand $E(t)$, the data center has to make an irrevocable electricity purchase decision on $\{R_e(t),R_b(t),G_e(t),G_b(t),B_e(t)\}$ in real time without knowing exactly these future information.

\section{Online Algorithm}

\subsection{Problem Relaxation}

In \textbf{P1}, the energy demand $E(t)$, electricity prices $P(t)$, and renewable availability $R(t)$ are not known in advance until time $t$. Therefore, solving the problem directly is impractical. Meanwhile, we can observe that the decision variables $G_b(t)$, $G_e(t)$, and $B_e(t)$ are subject to constraints imposed by the battery's SoC at each time $t$, which results in the problem being temporally coupled and challenging to resolve. To address the above issues, we first relax it to ensure constraint satisfaction over the long term. Define the time-average rate of charge and discharge as follows:
\begin{equation}\label{Gb}
    \overline{G_b} = \lim \limits_{T \to \infty} \frac{1}{T}\sum_{t=0}^{T-1} G_b(t) 
\end{equation}
\begin{equation}\label{Rb}
    \overline{R_b} = \lim \limits_{T \to \infty} \frac{1}{T}\sum_{t=0}^{T-1}  R_b(t) 
\end{equation}
\begin{equation}\label{Be}
    \overline{B_e} = \lim \limits_{T \to \infty} \frac{1}{T}\sum_{t=0}^{T-1}  B_e(t) .
\end{equation}
Therefore, we have $\overline{G_b}+\overline{R_b} = \overline{B_e}$. This is because if our charging/discharging consistently satisfies the constraints of the original problem, it will also satisfy such equations in the long term. We first relax the problem \textbf{P1} to obtain the problem \textbf{P2}:
\begin{equation} \label{P2}
    \textbf{P2: }\min \lim \limits_{T \to \infty} \frac{1}{T}\sum_{t=0}^{T-1} \mathbb{E} \{C(t) \}
\end{equation}
\begin{center}
    \text{s.t.} \eqref{V1},\eqref{V2},\eqref{V3},\eqref{V7},\eqref{eq:renewable},\eqref{E1},\eqref{C2}
\end{center}
\begin{equation}\label{new cons}
   \overline{G_b}+\overline{R_b} = \overline{B_e}.
\end{equation}
It is clear that \textbf{P2} is the optimization problem obtained by relaxing the constraints of \textbf{P1}. Therefore, any solution satisfying the constraints of \textbf{P1} is feasible for \textbf{P2}. The optimal value of \textbf{P2} is guaranteed to be less than that of \textbf{P1}. Denote the optimal objective value of \textbf{P1} as $Y^{OPT}$ and the optimal objective value of \textbf{P2} as $Y^{REL}$, that is $Y^{REL}\leq Y^{OPT}$. Then we employ the Lyapunov optimization technique \cite{neely2010stochastic} to transform the offline optimization problem into an online optimization problem that is independent of the statistical properties of $E(t)$, $R(t)$, and $P(t)$, thereby obtaining an approximate solution to the original problem. 

\subsection{Decoupling and Real-Time Decision via Lyapunov Optimization}
We employ the idea of Lyapunov optimization to enable real-time decisions. 
We define a virtual queue based on the SoC $B(t)$ of the virtual battery as 
\begin{equation}
    Q(t) = B(t) -\bar{B}_{min} - VP_{max} -  B_{dis},
\end{equation}
which is essentially a constant shift from $B(t)$ and where $B_{dis}:= \max_t B_{dis}(t)$ and $\bar{B}_{min} := \max_t B_{min}(t)$. Therefore, the virtual queue involves according to
\begin{equation}\label{queue}
    Q(t+1) = Q(t)+ G_b(t)+R_b(t)- B_e(t) .
\end{equation}
The virtual queue can decompose the long-term constraints in problem \textbf{P2} into each time $t$, thus removing the long-term constraints in \textbf{P2}. 
By setting specific relationships and weight parameter $V$, we can ensure that real-time decisions based on Lyapunov optimization satisfy all constraints of \textbf{P1}.

Before introducing our online algorithm, we design the Lyapunov function as $L(Q(t)) = \frac{1}{2} Q^2(t)$ and the conditional one-slot Lyapunov drift as follows:
\begin{equation}
    \Delta (Q(t)) = \mathbb{E}\{L(Q(t+1))-L(Q(t))|Q(t)\} .
\end{equation}
Note that taking square of both sides of \eqref{queue} yields
\begin{equation*}
\begin{aligned}
    Q^2(t+1) = Q^2(t) + [G_b(t)+R_b(t)-B_e(t)]^2 \\+2Q(t)[G_b(t)+R_b(t)-B_e(t)].
\end{aligned}
\end{equation*} 
Due to (\ref{V1}), (\ref{V3}) and (\ref{V7}), we have 
\begin{equation*}
\begin{aligned}
    & [G_b(t)+R_b(t)-B_e(t)]^2 \\
       =  &[G_b(t)+R_b(t)]^2 + B_e(t)^2 - 2[G_b(t)+R_b(t)]B_e(t)\\
    \leq  &[G_b(t)+R_b(t)]^2 + B_e(t)^2\\
    \leq  &\max\{B^2_{char}(t), B^2_{dis}(t)\}\\
    \leq  &\max\{B^2_{char}, B^2_{dis}\}
\end{aligned}
\end{equation*}
with $B_{char}:=\max_t B_{char}(t)$ and $B_{dis}:=\max_t B_{dis}(t)$. 
Define $B := \frac{1}{2} \max\{B^2_{char}, B^2_{dis}\}$. We can further have
\begin{equation*}
\begin{aligned}
    \frac{1}{2} (Q^2(t+1)-Q^2(t)) \leq B +  Q(t)[G_b(t)+R_b(t)-B_e(t)].
\end{aligned}
\end{equation*}
Taking the conditional expectation of both sides gives the following bound on the one-step Lyapunov drift:
\begin{equation}\label{Lya drift}
     \Delta(Q(t))
     \leq B + Q(t)\mathbb{E}\{G_b(t)+R_b(t)-B_e(t)|Q(t)\} .
\end{equation} 



By minimizing the Lyapunov drift plus the energy cost, we can lower the energy cost and simultaneously ensure the stability of the virtual queue. Instead of directly controlling the exact Lyapunov drift, we take a conservative alternative and limit its upper bound, i.e., the right-hand side of (\ref{Lya drift}). 
To this end, we develop an online algorithm that decides $\{R_e(t),R_b(t),G_e(t),G_b(t),B_e(t)\}$ in real time based on the observation of $\{R(t), P(t), E(t)\}$ by solving 
\begin{equation}\label{real time contrl}
\begin{aligned}
  \textbf{P3: }\min & [Q(t)+VP(t)]G_b(t) + Q(t)R_b(t) \\
    & - [Q(t) +VP(t)]B_e(t) -VP(t)R_e(t).
\end{aligned}
\end{equation}
\begin{center}
    \text{s.t.} \eqref{V1},\eqref{V2},\eqref{V3},\eqref{V7},\eqref{eq:renewable},\eqref{E1},\eqref{C2}
\end{center}
where the objective reflects the weighted sum of the upper bound on the Lyapunov drift and the energy cost (with a constant shift). We summarize the proposed online algorithm in Algorithm 1:

\begin{algorithm}\label{online alg} %
    \caption{Online Algorithm}
        At each time $t$:\\
        1. Observe $R(t), P(t), E(t)$ and $Q(t)$;\\
        2.  Determine $\left(R^*_e(t),R^*_b(t),G^*_e(t),G^*_b(t),B^*_e(t)\right)$ by solving \textbf{P3};\\
        3. Implement the decision and update $B(t+1)$ following the dynamics (\ref{V4}).
\end{algorithm}

\subsection{Performance Analysis}
Before summarizing the properties of the algorithm, we first analyze the optimal solution of \textbf{P3}.
\begin{lemma}\label{lemma1} 
The optimal solution to \textbf{P3} satisfies the following:
\begin{enumerate}
    \item If  $Q(t)+VP(t) \leq 0$ holds, we have $B^*_e(t) = 0$.
    \item If  $Q(t) > 0$ holds, we have $G^*_b(t) = 0$ and $R^*_b(t) = 0$.
\end{enumerate}
\end{lemma}
See Appendix A for the proof. Lemma~\ref{lemma1} contributes to more insights into Algorithm~1 as follows.
\begin{theorem}\label{theorem1} 
If $R(t)$, $E(t)$, $P(t)$, $\forall t$, are i.i.d., Algorithm~1 achieves a time-average energy cost more than the optimum $Y^{OPT}$ by at most a constant $B/V$, i.e.,
\begin{equation*}
    \lim \limits_{T \to \infty} \frac{1}{T}\sum_{t=0}^{T-1} \mathbb{E} \{C(t)\} \leq Y^{OPT}+B/V.
\end{equation*}
Moreover, given $B_{char}:= \max_t B_{char}(t)$ and $\bar{B}_{max}:=\min_t B_{max}(t)$, 
if $0 < V \leq V_{max} $ holds with
\begin{equation}
    V_{max} := \frac{\bar{B}_{max}- \bar{B}_{min}- B_{dis}- B_{char}}{P_{max}},
\end{equation}
it is guaranteed that 
\begin{equation*}
  {B}_{min}(t)  \le B(t) \le {B}_{max}(t), \quad \forall t,
\end{equation*}
holds.


\end{theorem}
See Appendix B for the proof.

\begin{remark}
    In {Theorem \ref{theorem1}}, as $V$ increases, the long-term average energy cost will get closer to the optimal value of the original problem \textbf{P1}. 
    However, there exists an upper bound for the choice of $V$ in order to guarantee the feasibility of virtual battery operation. It can always be achieved by setting a sufficiently small $V$.
    The tradeoff implies that a good estimate of $V_{max}$ is necessary.
    In practice, even if we set $V> V_{max}$,
    we can project any infeasible real-time decision onto the feasible region of the virtual battery given the latest observation.
\end{remark}

\section{Numerical Results}
In this section, we evaluate the proposed algorithm through simulation experiments. We consider a total time duration of 30 days, where each time $t$ corresponds to 1 hour. 
The electricity prices $P(t)$, fluctuate randomly within the range [0.5, 1.5]. The electricity price is in the unit of \textdollar/kWh.
The hourly energy requirement $E(t)$ fluctuates within the interval [10000, 20000] in terms of kWh.
The renewable energy generation $R(t)$ fluctuates hourly within the range of [0, 3000] kWh.
The maximum capacity $B_{max}(t)$ of the aggregated virtual battery fluctuates randomly within the range of 3000 kWh to 4000 kWh, while its minimum capacity $B_{min}(t)$ varies randomly within the range of 1000 kWh to 2000 kWh. The charging rate $B_{char}(t)$ and discharging rate $B_{dis}(t)$  can fluctuate arbitrarily between 100 kW and 200 kW. We assume that we know $\Bar{B}_{max}=3000,\Bar{B}_{min} = 2000, B_{char} = 200, B_{dis} = 200$ in advance. Hence we have $V_{max} = 400$.

\begin{figure}[htbp]
    \centering
    \includegraphics[scale=0.35]{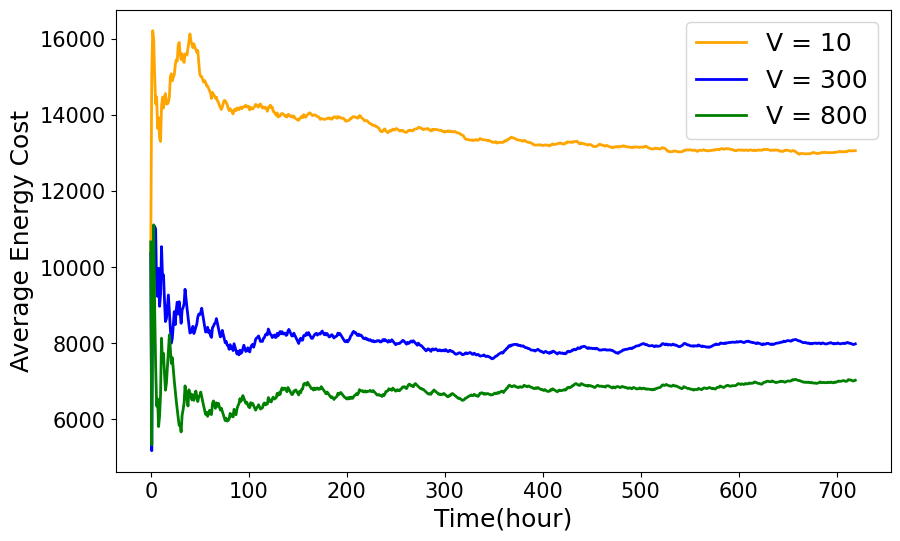}
    \caption{Average energy cost with different $V$}
    \label{f2}
\end{figure}

\textbf{Impact of parameter $V$}. First, we observe how the parameter $V$ inﬂuences the energy cost reduction. It is seen in Fig.\ref{f2} that the average cost decreases with increasing $V$. The results align with our algorithmic performance demonstrated in \textbf{Theorem \ref{theorem1}}. As intuitively expected, our algorithm discharges from the virtual battery during periods of high electricity prices and charges it during periods of low electricity prices. Additionally, it stores excess renewable energy for subsequent use when electricity prices become excessively high.

\begin{figure}[htbp]
    \centering
    \includegraphics[scale=0.35]{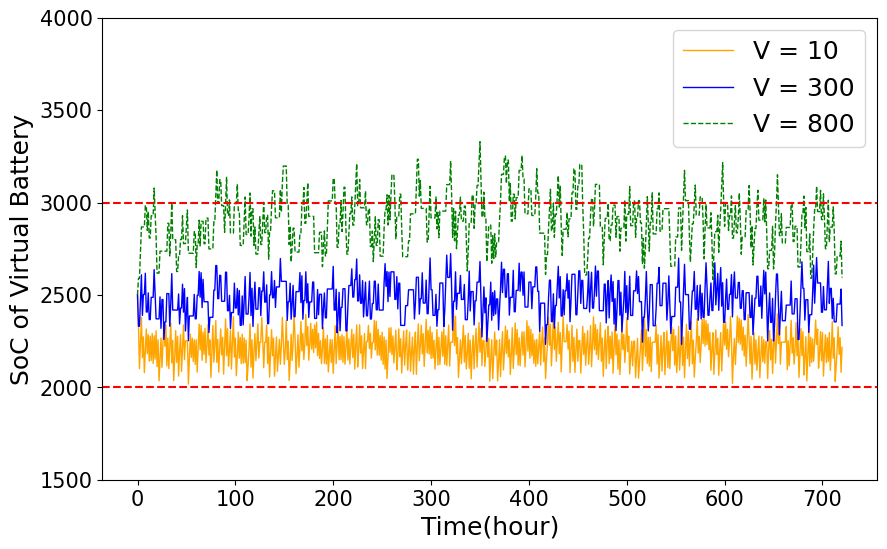}
    \caption{Variation of SoC under different settings of $V$}
    \label{f3}
\end{figure}

\textbf{Variation of virtual battery SoC}. Then, we observe whether the virtual battery always remains between its upper and lower bounds under our algorithm, and under what circumstances it may exceed these bounds. When we know in advance $\Bar{B}_{max}, \bar{B}_{min}, B_{dis}, B_{char}$, and $P_{max}$, we can compute the upper limit of $V$, and set $V$ to its maximum attainable value. In this scenario, the virtual battery remains within its SoC bounds at all times. However, without prior knowledge of these parameters, we may not accurately determine the value of $V_{max}$. When we do not know the upper bound of $V$, inappropriate selection of $V$ may result in the SoC of the virtual battery exceeding its bounds. As illustrated in Fig.\ref{f3}, when we set $V = 10$ and $V = 300$, that is $V \leq V_{max}$, we observe that the SoC of the virtual battery remains within its bounds. However, when $V = 800 > V_{max}$ it can be observed that the SoC of the virtual battery may exceed its limits. Therefore, by setting an appropriate $V$, we can achieve bounded loss compared with the offline optimal cost.  

\section{Conclusions}
In this work, we propose a two-layer hierarchy where the lower layer consists of the operation of all electrical equipment in the data center and the upper layer determines the procurement and dispatch of electricity to minimize the long-term time-averaged energy cost. We further design an online algorithm based on Lyapunov optimization which can assist data centers in purchasing electricity from the grid under the uncertainties and fluctuations in energy prices, renewable energy availability, and battery specifications. With the increasing parameter $V$, our algorithm can achieve a cost that is close to the offline optimal. Simulation results validate the theoretical analysis and show the effectiveness of our algorithm in reducing the long-term cost. In future work, we will consider a more realistic virtual battery model with dissipation.


\bibliographystyle{unsrt}
\bibliography{reference}

\begin{thebibliography}{10}

\bibitem{zhou2017optimal}
Yaqin Zhou, David~KY Yau, Pengcheng You, and Peng Cheng.
\newblock Optimal-cost scheduling of electrical vehicle charging under uncertainty.
\newblock {\em IEEE Transactions on Smart Grid}, 9(5):4547--4554, 2017.

\bibitem{10136373}
Zhe Wang, Heng Zhang, Xianghui Cao, Endong Liu, Hongran Li, and Jian Zhang.
\newblock Modeling and detection scheme for zero-dynamics attack on wind power system.
\newblock {\em IEEE Transactions on Smart Grid}, 15(1):934--943, 2024.

\bibitem{toosi2015fuzzy}
Adel~Nadjaran Toosi and Rajkumar Buyya.
\newblock A fuzzy logic-based controller for cost and energy efficient load balancing in geo-distributed data centers.
\newblock In {\em 8th International Conference on Utility and Cloud Computing}, pages 186--194. IEEE, 2015.

\bibitem{chen2017stochastic}
Tianyi Chen, Aryan Mokhtari, Xin Wang, Alejandro Ribeiro, and Georgios~B Giannakis.
\newblock Stochastic averaging for constrained optimization with application to online resource allocation.
\newblock {\em IEEE Transactions on Signal Processing}, 65(12):3078--3093, 2017.

\bibitem{neely2010efficient}
Michael~J Neely, Arash~Saber Tehrani, and Alexandros~G Dimakis.
\newblock Efficient algorithms for renewable energy allocation to delay tolerant consumers.
\newblock In {\em 1st International Conference on Smart Grid Communications}, pages 549--554. IEEE, 2010.

\bibitem{wei2017data}
Xiaohan Wei and Michael~J Neely.
\newblock Data center server provision: Distributed asynchronous control for coupled renewal systems.
\newblock {\em IEEE/ACM Transactions on Networking}, 25(4):2180--2194, 2017.

\bibitem{yao2012power}
Yuan Yao, Longbo Huang, Abhishek~B Sharma, Leana Golubchik, and Michael~J Neely.
\newblock Power cost reduction in distributed data centers: A two-time-scale approach for delay tolerant workloads.
\newblock {\em IEEE Transactions on Parallel and Distributed Systems}, 25(1):200--211, 2012.

\bibitem{zhang2020distributed}
Guanglin Zhang, Shun Zhang, Wenqian Zhang, Zhirong Shen, and Lin Wang.
\newblock Distributed energy management for multiple data centers with renewable resources and energy storages.
\newblock {\em IEEE Transactions on Cloud Computing}, 10(4):2469--2480, 2020.

\bibitem{nayyar2013aggregate}
Ashutosh Nayyar, Josh Taylor, Anand Subramanian, Kameshwar Poolla, and Pravin Varaiya.
\newblock Aggregate flexibility of a collection of loads.
\newblock In {\em 52nd Conference on Decision and Control}, pages 5600--5607. IEEE, 2013.

\bibitem{hao2014aggregate}
He~Hao, Borhan~M Sanandaji, Kameshwar Poolla, and Tyrone~L Vincent.
\newblock Aggregate flexibility of thermostatically controlled loads.
\newblock {\em IEEE Transactions on Power Systems}, 30(1):189--198, 2014.

\bibitem{hao2014characterizing}
He~Hao and Wei Chen.
\newblock Characterizing flexibility of an aggregation of deferrable loads.
\newblock In {\em 53rd Conference on Decision and Control}, pages 4059--4064. IEEE, 2014.

\bibitem{hao2015generalized}
He~Hao, Abhishek Somani, Jianming Lian, and Thomas~E Carroll.
\newblock Generalized aggregation and coordination of residential loads in a smart community.
\newblock In {\em 6th International Conference on Smart Grid Communications}, pages 67--72. IEEE, 2015.

\bibitem{madjidian2017energy}
Daria Madjidian, Mardavij Roozbehani, and Munther~A Dahleh.
\newblock Energy storage from aggregate deferrable demand: Fundamental trade-offs and scheduling policies.
\newblock {\em IEEE Transactions on Power Systems}, 33(4):3573--3586, 2017.

\bibitem{zhao2017geometric}
Lin Zhao, Wei Zhang, He~Hao, and Karanjit Kalsi.
\newblock A geometric approach to aggregate flexibility modeling of thermostatically controlled loads.
\newblock {\em IEEE Transactions on Power Systems}, 32(6):4721--4731, 2017.

\bibitem{taha2022efficient}
Feras~Al Taha, Tyrone Vincent, and Eilyan Bitar.
\newblock An efficient method for quantifying the aggregate flexibility of plug-in electric vehicle populations.
\newblock {\em arXiv preprint arXiv:2207.07067}, 2022.

\bibitem{fang_thermal-aware_2017}
Qiu Fang, Jun Wang, Qi~Gong, and Mengxuan Song.
\newblock Thermal-aware energy management of an hpc data center via two-time-scale control.
\newblock {\em IEEE Transactions on Industrial Informatics}, 13(5):2260--2269, 2017.

\bibitem{neely2010stochastic}
Michael Neely.
\newblock {\em Stochastic network optimization with application to communication and queueing systems}.
\newblock Springer Nature, 2022.

\bibitem{huang2011utility}
Longbo Huang and Michael~J Neely.
\newblock Utility optimal scheduling in processing networks.
\newblock {\em Performance Evaluation}, 68(11):1002--1021, 2011.

\end{thebibliography}

\section*{Appendix}

\subsection{Proof of Lemma \ref{lemma1}}
In this part, we show the solution in some cases below:

1. $For \quad Q(t) < Q(t) +VP(t) \leq 0$: 
To minimize the objective of \textbf{P3}, we strive to maximize the values of $G_b(t)$, $R_b(t)$, and $R_e(t)$ while minimizing $B_e(t)$ as much as possible. Since $Q(t) \leq -VP(t)$ and $Q(t) < Q(t) + VP(t)$, maximizing $R_b(t)$ can further reduce the value of \textbf{P3}. Thus in this case we can obtain:
\begin{equation*}
\left\{
\begin{aligned}
    & R_b(t) = \min \{R(t), B_{char}(t)\}\\
    & B_e(t) = 0 \\
    & R_e(t) = \min\{R(t) - R_b(t), E(t) - B_e(t)\} \\
    & G_b(t) = B_{char}(t) - R_b(t) \\
    & G_e(t) = E(t) - R_e(t) - B_e(t).
\end{aligned}
\right.  
\end{equation*}


2. $For \quad Q(t) \leq 0 < Q(t)+VP(t) $: To minimize the objective of \textbf{P3}, we need the solution is to set $R_b(t)$, $B_e(t)$ and $R_e(t)$ as large as possible and $G_b(t)$ as small as possible. Since $ Q(t)+VP(t) > 0$ and $Q(t) \leq 0$ , we can conclude $-VP(t) <  Q(t)$ and $-VP(t) \leq -(Q(t) + VP(t))$. This means maximizing $R_e(t)$ in the can further reduce the value of \textbf{P3}. Because the virtual battery cannot charge and discharge simultaneously, one possible solution when the virtual battery discharges is:
\begin{equation*}
\left\{
\begin{aligned}
    & R_e(t) = \min \{R(t), E(t)\}\\
    & B_e(t) = \min\{B_{dis}(t), E(t) - R_e(t)\}  \\
    & R_b(t) = 0 \\
    & G_b(t) = 0 \\
    & G_e(t) = E(t) - R_e(t) - B_e(t).
\end{aligned}
\right.  
\end{equation*}
There exists another solution which is to keep the virtual battery charging as follows:
\begin{equation*}
\left\{
\begin{aligned}
    & R_e(t) = \min \{R(t), E(t)\}\\
    & B_e(t) = 0  \\
    & R_b(t) = \min\{R(t) - R_e(t), B_{char}(t)\} \\
    & G_b(t) = 0 \\
    & G_e(t) = E(t) - R_e(t) - B_e(t).
\end{aligned}
\right.  
\end{equation*}
3. $For \quad 0 < Q(t)  < Q(t)+VP(t)$: To minimize the objective of \textbf{P3}, we need to make the values of $G_b(t)$, $R_b(t)$ as small as possible, while maximizing $R_e(t)$ , $B_e(t)$ as much as possible. Thus we can obtain the following:
\begin{equation*}
\left\{
\begin{aligned}
    & G_b(t) = 0\\
    & R_b(t) = 0 \\
    & B_e(t) = \min\{B_{dis}(t), E(t)\}\\
    & R_e(t) = \min\{R(t)-R_b(t), E(t) - B_e(t) \} \\
    & G_e(t) = E(t) - B_e(t) - R_e(t).
\end{aligned}
\right.  
\end{equation*}

\subsection{Proof of Theorem \ref{theorem1}}
Before proving this part, we need to use the following lemma.
\begin{lemma}
If the $R(t)$, $E(t)$, and $P(t)$ are i.i.d. over time $t$, then there exists a stationary, randomized policy that satisfies the constraints \eqref{V1},\eqref{V2},\eqref{V3},\eqref{V7},\eqref{eq:renewable},\eqref{E1},\eqref{C2} and provides the following guarantees:
\begin{equation}
    \mathbb{E}\{  (G^*_b(t)+R^*_b(t))\} = \mathbb{E}\{  B^*_e(t)\}
\end{equation}
\begin{equation}
    \mathbb{E}\{P(t)(G^*_b(t)+G^*_e(t))\} = Y^{REL}
\end{equation}
where the expectations above are with respect to the stationary distribution of $\{E(t), R(t), P(t)\}$ and the randomized control decisions. 
\end{lemma}
This result has been proven in \cite{huang2011utility}. It is omitted here for brevity.

Because $ \Delta(Q(t)) \leq B + Q(t)\mathbb{E}\{  (G_b(t)+R_b(t))- B_e(t)|Q(t)\} $, 
we have $ \Delta(Q(t)) + V\mathbb{E}\{P(t)(G_b(t)+G_e(t))|Q(t)\} \leq B + Q(t)\mathbb{E}\{  (G_b(t)+R_b(t))-  B_e(t)|Q(t)\}+V\mathbb{E}\{P(t)(G_b(t)+G_e(t))|Q(t)\} $, we could see our online algorithm is designed to minimize the right-hand side of above inequality over all possible feasible control decisions. This includes the optimal, stationary, randomized policy given in \textbf{Lemma 2}. Therefore, we can obtain the following inequality:
\begin{equation}\label{proof3}
\begin{aligned}
    &\Delta(Q(t)) + V\mathbb{E}\{P(t)(G_b(t)+G_e(t))|Q(t)\} \\
    \leq& B + Q(t)\mathbb{E}\{  (G^*_b(t)+R^*_b(t))-  B^*_e(t)|Q(t)\}\\
    &+V\mathbb{E}\{P(t)(G^*_b(t)+G^*_e(t))|Q(t)\}\\
    \leq & B + VY^{REL} \leq B + VY^{OPT}.
\end{aligned}
\end{equation}
By taking the expectation of both sides of (\ref{proof3}) and summing over $t \in \{0,1,2,...,T-1\}$, we can get
\begin{equation}
\begin{aligned}
    &V\sum_{t=0}^{T-1}\mathbb{E}\{C(t)\} \\
    &\leq TB+TVY^{OPT} - \mathbb{E}\{L(Q(T)\} + \mathbb{E}\{L(Q(0)\} , 
\end{aligned}
\end{equation}
where we have applied Law of Total Expectation.
Then we divide both sides by $V\cdot T$. When $T \rightarrow \infty$, owing to that $\mathbb{E}\{L(Q(T)\}$ and $\mathbb{E}\{L(Q(0)\} $ are finite, we can get:
\begin{equation}
     \lim \frac{1}{T}\sum_{t=0}^{T-1}\mathbb{E}\{C(t)\} \leq Y^{OPT} + \frac{B}{V} .
\end{equation}
Next, we prove the feasibility of virtual battery operation when $V$ is properly chosen. For a specific $t$ and $\bar{B}_{min} \leq B(t) \leq \bar{B}_{max}$, we have ${-VP_{max}}- B_{dis} \leq Q(t) \leq \bar{B}_{max} -\bar{B}_{min} - VP_{max}- B_{dis}$.
Now assuming that the time $t$ complies with the aforementioned bounds, we need to prove that it also holds for time $t+1$.

First suppose $ -VP_{max}-B_{dis} \leq  Q(t) \leq -VP_{max} $. In this case, by the \textbf{Lemma 1}, it can be inferred that $B_e(t) = 0$, then
\begin{equation*}
\begin{aligned}
    Q(t+1) 
    & \geq -VP_{max} -   B_{dis} +   [G_b(t)+R_b(t)]\\
    & \quad -   B_e(t)\\
    &\geq -VP_{max}-  B_{dis}
\end{aligned}
\end{equation*}
\begin{equation*}
\begin{aligned}
    Q(t+1) &\leq -VP_{max}  +   [G_b(t)+R_b(t)]-  B_e(t)\\
    &\leq    B_{char}\\
    &\leq \bar{B}_{max} -\bar{B}_{min}-  B_{dis}-VP_{max}.
\end{aligned}
\end{equation*}
In the proof of the upper bound, we utilized $ 0 < V \leq V_{max}$.

Second, suppose $-VP_{max} <  Q(t) \leq  0 $, then
\begin{equation*}
\begin{aligned}
    Q(t+1) & > -VP_{max} +   [G_b(t)+R_b(t)]-  B_e(t)\\
    & > -VP_{max}-  B_e(t)\\
    & > -VP_{max}-  B_{dis}
\end{aligned}
\end{equation*}
\begin{equation*}
\begin{aligned}
    Q(t+1) &\leq   [G_b(t)+R_b(t)]-  B_e(t)\\
    & \leq    B_{char}\\
    &\leq  \bar{B}_{max} -\bar{B}_{min} -  B_{dis}-VP_{max}.
\end{aligned}
\end{equation*}
In proving the upper bound, we also utilize $ 0 < V \leq V_{max}$.

Finally, suppose $0 <  Q(t) \leq  \bar{B}_{max} -\bar{B}_{min}-VP_{max} -  B_{dis}$. In this case, by the \textbf{Lemma 1} we can infer $G_b(t) = 0$ and $R_b(t) = 0$, then
\begin{equation*}
\begin{aligned}
    Q(t+1) &>   [G_b(t)+R_b(t)]-  B_e(t) \\
    & > -  B_{dis}\\
    & >  -VP_{max}-  B_{dis}
\end{aligned}
\end{equation*}
\begin{equation*}
\begin{aligned}
    Q(t+1) &\leq \bar{B}_{max} -\bar{B}_{min}-  B_{dis}\\
    &\quad -VP_{max}+  [G_b(t)+R_b(t)]-  B_e(t)\\
    &\leq \bar{B}_{max} -\bar{B}_{min}-VP_{max}-  B_{dis}.
\end{aligned}
\end{equation*}
Through the aforementioned induction, we have established the upper and lower bounds of the $Q(t)$.

Since $Q(t)$ always remains between the upper and lower bounds and $Q(t)= B(t) -\bar{B}_{min} - VP_{max}-  B_{dis}$, then we have 
$B(t) -\bar{B}_{min} -VP_{max}-  B_{dis} \geq -VP_{max}-  B_{dis}$ and $ B(t) -\bar{B}_{min} -VP_{max}-  B_{dis} \leq B_{max}^{max}-\bar{B}_{min}-  B_{dis}-VP_{max}$. Therefore in each $t$, we conclude the following holds:
\begin{equation*}
    B_{min}(t) \leq \bar{B}_{min} \leq B(t) \leq \bar{B}_{max} \leq B_{max}(t) .
\end{equation*}
Our algorithm satisfies the battery constraints at each time $t$, thereby ensuring that all decisions are feasible and do not violate constraints.

\end{document}